\documentclass{article}
\usepackage{spconf,amsmath,graphicx,bm,amssymb,cite,flushend}
\usepackage[font=footnotesize]{subfig}

\newtheorem{theorem}{Theorem}
\newtheorem{remark}{Remark}

\newtheorem{definition}{Definition}
\newtheorem{example}{Example}
\usepackage{flushend}
\usepackage{color}

\usepackage{algorithm}
\usepackage{algorithmic}

\title{Atomic Norm Based Localization of Far-Field and Near-Field Signals with Generalized Symmetric Arrays}
\name{Xiaohuan Wu$^1$, Wei-Ping Zhu$^{1, 2}$ and Jun Yan$^1$\thanks{This work was supported by the National Natural Science Foundation of China (61801245, 61772287 and 61771256), the NSF of Jiangsu Province (BK20180748), the NSF of Jiangsu Higher Education Institutions (18KJB510032), NUPTSF (NY218101), open research fund of Key Lab of Broadband Wireless Communication and Sensor Network Technology (JZNY201913) and the Key University Science Research Project of Jiangsu Province (18KJA510004).}}
\address{$^1$School of Telecommunication and Information Engineering, \\
        Nanjing University of Posts and Telecommunications, Nanjing, China\\
        $^2$Department of Electrical and Computer Engineering,\\ Concordia University, Montreal, Canada\\ Email: \{xiaohuanwu, zwp, yanj\}@njupt.edu.cn}
\begin{document}
%
\maketitle
\begin{abstract}
Most localization methods for mixed far-field (FF) and near-field (NF) sources are based on uniform linear array (ULA) rather than sparse linear array (SLA). In this paper, we propose a localization method for mixed FF and NF sources based on the generalized symmetric linear arrays, which include ULAs, Cantor array, Fractal array and many other SLAs. Our method consists of two steps. In the first step, the high-order statistics of the array output is exploited to increase the degree of freedom. Then the direction-of-arrivals (DOAs) of the FF and NF sources are jointly estimated by using the recently proposed atomic norm minimization (ANM), which belongs to the gridless super-resolution method since the discretization of the parameter space is not required. In the second step, the ranges are given by MUSIC-like one-dimensional searching.
Simulations results are provided to demonstrate the advantages of our method.
\end{abstract}
\begin{keywords}
Source localization, far-field, near-field, generalized symmetric arrays, atomic norm minimization.
\end{keywords}

\section{Introduction}
\label{sec Introduction}

Source localization is a fundamental problem in array signal processing and has received considerable attention. Based on the far-field (FF) source assumption where the impinged signals are assumed to be plane-wave, numerous methods have been proposed for FF source localization, i.e., direction-of-arrival (DOA) estimation \cite{MUSIC1986schmidt,esprit1989roy,malioutov2005sparse,zhouchengwei2017TVT,my2018TVT}\cite{Zhouchengwei2017SJ}. However, when the sources are close to the array and lie in the near-field (NF) region (i.e., the Fresnel region), the impinged signals are spherical wave rather than plane-wave. In this case, the steering vectors are characterized by two independent parameters: DOA and range. Hence, the DOA estimation methods for FF source localization can not be applied to the NF source localization.
In order to deal with this problem, the nonlinear time delay of the spherical wavefront model is approximated into a quadratic wavefront model by using the second-order Taylor expansion. Based on this approximation, a large number of NF source localization methods have been proposed. For instance, He \emph{et al.} have proposed an oblique projection based MUSIC (OPMUSIC) algorithm for mixed FF and NF sources localization \cite{He2012TSP_OPMUSIC}. An efficient subspace-based localization method without eigendecomposition has been proposed in \cite{zuo2018TSP}. However, the maximum number of resolvable sources of these methods are limited to half of the number of sensors. Liang \emph{et al.} have proposed a cumulant based algorithm called two-stage MUSIC (TSMUSIC) by using the high-order statistics (HOS) of the array output \cite{liang2010TSP_TSMUSIC}. By using the HOS and compressed sensing theory, a sparse method has been presented for mixed sources localization \cite{Wang2012SPL_sparseNF}. However, sparse method requires to discretize the angle space hence its performance is limited by the so called "basis mismatch" effect. Note that, all the aforementioned methods are proposed based on the uniform linear array (ULA) and most of them can not be extended to the sparse linear array (SLA). As a result, the maximum number of resolvable sources of these methods can not exceed the number of sensors.

Recently, the SLA has been studied in FF DOA estimation by exploiting the coarray of the physical array to increase the array aperture \cite{PPal2011TSP_coprime, PPal2011TSP_Nested,zhouchengwei2013WCSP,zhouchengwei2017IETcom}. Due to the extended array aperture, the resolution and the maximum number of resolvable sources can be greatly improved. Nevertheless, the classical SLAs, e.g., coprime array, nested array can not be directly employed for NF source localization since most of the NF source localization methods require symmetric array structure. By using the special geometry of the nested array (i.e., two subarrays without overlapping), several symmetric nested arrays have been proposed \cite{Wang2013SPL_nested, Li2016Sensors_nested}. However, these literatures focus on constructing specific SLA to extend the array aperture. A unified algorithm for generalized symmetric SLA in NF source localization area is still missing. Furthermore, all these methods choose MUSIC for DOA estimation which suffers from basis mismatch effect. 

In this paper, we propose a unified localization method for mixed FF and NF sources which can be applied to any sparse/uniform symmetric redundancy array. The high-order cumulants of the array output are exploited to increase the degrees-of-freedom (DoFs). The atomic norm is employed to eliminate the basis mismatch effect in DOA estimation and able to provide super-resolution. Moreover, the proposed method does not require the prior knowledge of the noise power as well as the number of sources or FF/NF sources.

\section{Signal Model}
\label{sec signal model}

We first give the definition of the generalized symmetric linear array that will be used throughout of this paper. Let $[N] = \{-N,\cdots,-1,0,1,\cdots,N\}$. The indices of the array sensors are denoted as $\bm{\Omega} = \{ \Omega_{-M},\cdots, \Omega_{-1}, \Omega_0,\Omega_1,\cdots,\Omega_M \} \subseteq [N]$ where $\{\Omega_0,\cdots,\Omega_M\}$ are non-negative unequal integers and sorted ascendingly with $\Omega_0 = 0$ and $\Omega_M = N$. Due to the symmetric property of the array, we have $\Omega_{-m} = -\Omega_m, m\in[M]$.
Let $|\bm{\Omega}|$ denote the cardinality of $\bm{\Omega}$, it is easy to see that $M = \frac{|\bm{\Omega}|-1}{2} \leq N$. If $N = M$, $\bm{\Omega}$ represents a symmetric ULA while if $N>M$, $\bm{\Omega}$ represents a symmetric SLA. It should be noted that ULA can be regarded as a special case of SLA. As a result, only the SLA case is considered in the rest of this paper. We then provide the definition of coarray below.
\begin{definition}
  The coarray $\mathcal{D}$ of a physical array $\bm{\Omega}$ is defined as, $\mathcal{D} = \{\Omega_m - \Omega_n - N: m,n\in \bm{\Omega}, m\geq n\} \subseteq [N]$. $\bm{\Omega}$ is called a redundancy array if $\mathcal{D} = [N]$ (a.k.a. hole-free coarray). Otherwise, $\bm{\Omega}$ is called a non-redundancy array.
\end{definition}
It is shown that the maximum number of detectable sources using the array $\bm{\Omega}$ is determined by its coarray $\mathcal{D}$ rather than the number of physical sensors \cite{zhouchengwei2018SPL}. In fact, a redundancy array is able to detect up to $2N$ sources, which is greater than its number of sensors $2M+1$ if $N>M$. To better understand our array, we here provide a simple example.
\begin{example}\label{exa SLA}
  A symmetric SLA with $\bm{\Omega} = \{-3, -2, 0, 2, 3\}$ has the coarray $\mathcal{D} = [3]$ which is hole-free. By using proper algorithms, the array can detect up to $2N = 6$ sources, which is greater than the number of sensors.
\end{example}
It should be pointed out that the aforementioned symmetric SLAs with hole-free coarray not only include the array in Example \ref{exa SLA}, but also contain the recently proposed cantor array \cite{Liuchunlin2017CAMSAP_CantorArray} and fractal array \cite{Eldar2019ICASSP_FractalArray}.\footnote{Note that the factal array is able to extend any known symmetric redundancy array to a much larger one. }. In fact, the proposed method can be applied to any symmetric redundancy array. To avoid the angle ambiguity, the minimum sensor spacing $d$ should satisfy $d \leq \lambda/4$ where $\lambda$ is the wavelength.


We then assume $K$ narrowband uncorrelated mixed FF and NF sources impinge onto the array and the $k$-th source is characterized by a parameter pair $(\theta_k,r_k)$.
The array output is given by,
\begin{equation}
  \bm{x}_{\bm{\Omega}}(t) = \bm{A}_{\bm{\Omega}}\bm{s}(t) + \bm{n}_{\bm{\Omega}}(t),
\end{equation}
where $\bm{x}_{\bm{\Omega}}(t) = [x_{\Omega_{-M}}(t),\cdots,x_{\Omega_M}(t)]^T$ is the array output, $\bm{s}(t) = [s_1(t),\cdots,s_K(t)]^T$ is the signal waveform, $\bm{n}_{\bm{\Omega}}(t)$ denotes the additive white Gaussian noise with zero mean and $\bm{A}_{\bm{\Omega}} = [\bm{a}_{\bm{\Omega}}(\theta_1,r_1),\cdots,\bm{a}_{\bm{\Omega}}(\theta_K,r_K)]$ denotes the manifold matrix of the array, in which $\bm{a}_{\bm{\Omega}}(\theta_k,r_k) = [ e^{j[-\Omega_M\omega_k+(-\Omega_M)^2\phi_k]}, \cdots, 1, \cdots, e^{j[\Omega_M\omega_k+\Omega_M^2\phi_k]} ]^T$ is the steering vector of the $k$-th signal with
\begin{equation}
  \omega_k = -2\pi \frac{d}{\lambda}\sin(\theta_k),
\end{equation}
\begin{equation}
  \phi_k = \pi \frac{d^2}{\lambda r_k} \cos^2(\theta_k).
\end{equation}
Note that we unify the steering vectors of the FF and NF sources as $\bm{a}_{\bm{\Omega}}(\theta_k,r_k)$ since the steering vector of the FF source can be represented as $\bm{a}_{\bm{\Omega}}(\theta_k,+\infty)$.

\section{The Proposed Method}

\subsection{DOA Estimation of FF and NF Sources}

Similar to TSMUSIC, the fourth-order cumulants are utilized to construct a Hermitian matrix without the range parameter for DOA estimation. Specifically, we define the fourth-order cumulant as
\begin{equation}\label{eq culmulant}
\begin{split}
  \bm{C}_{\bm{\Omega}}(\bar{m},\bar{n})  &= -cum\{ x_{\Omega_m}(t), x^*_{-{\Omega_m}}(t), x_{-{\Omega_n}}(t), x^*_{{\Omega_n}}(t) \} \\
  &= \sum_{k=1}^K -c_{s_k} e^{ j2(\Omega_m - \Omega_n)\omega_k},
\end{split}
\end{equation}
where $\bar{m} = m+M+1$, $\bar{n} = n+M+1$, $m,n \in [M]$ and $c_{s_k}<0$ is the fourth-order cumulant of $s_k(t)$.\footnote{Since many practical signals such as sinusoidal signal, amplitude shift keying (ASK) signal and phase shift keying (PSK) signal are sub-Gaussian process, we assume $s_k(t)$ is zero-mean stationary random processes with negative kurtosis, i.e., $c_{s_k}<0$.} $\bm{C}_{\bm{\Omega}}$ can be compactly written as the following Hermitian matrix,
\begin{equation}\label{eq cul_matrix}
\begin{split}
  \bm{C}_{\bm{\Omega}} &= \sum_{k=1}^K -c_{s_k} \bar{\bm{a}}_{\bm{\Omega}}(\theta_k) \bar{\bm{a}}_{\bm{\Omega}}^H(\theta_k) \\
  &= \bar{\bm{A}}_{\bm{\Omega}} \bm{C_s} \bar{\bm{A}}_{\bm{\Omega}}^H,
\end{split}
\end{equation}
where $\bm{C_s} = \text{diag}([-c_{s_1},\cdots,-c_{s_K}])$ contains the fourth-order cumulants of the signals, $\bar{\bm{A}}_{\bm{\Omega}} = [\bar{\bm{a}}_{\bm{\Omega}}(\theta_1),\cdots,\bar{\bm{a}}_{\bm{\Omega}}(\theta_K)]$ and $\bar{\bm{a}}_{\bm{\Omega}}(\theta_k) = [e^{j2\Omega_{-M}\omega_k}, \cdots,1,\cdots, e^{j2\Omega_M\omega_k} ]^T$.
It can be seen from (\ref{eq cul_matrix}) that, the range parameter is removed from $\bm{C}_{\bm{\Omega}}$. Hence matrix $\bm{C}_{\bm{\Omega}}$ can be regarded as the covariance matrix of a virtual array output with manifold matrix $\bar{\bm{A}}_{\bm{\Omega}}$ which is positive semidefinite (PSD). Although we can apply the subspace based methods such as MUSIC to $\bm{C}_{\bm{\Omega}}$ to estimate the DOAs, directly using MUSIC for DOA estimation can not utilize the coarray property to increase the DoFs. Below we use the atomic norm theory which can fully utilize the coarray to estimate the DOAs.

First, denote $\bar{\bm{a}}(\theta_k) = [e^{j2(-N)\omega_k}, \cdots,1,\cdots, e^{j2N\omega_k} ]^T$ as the steering vector of the coarray $\mathcal{D}$ and $\bm{\Gamma}$ as a selecting matrix such that the $j$-th row of $\bm{\Gamma}$ contains all $0$s but a single $1$ at the $(\Omega_{j-M-1}+N+1)$-th position. It is clear that $\bar{\bm{a}}_{\bm{\Omega}}(\theta_k) = \bm{\Gamma}\bar{\bm{a}}(\theta_k)$ and
\begin{equation}\label{eq Aomega}
  \bar{\bm{A}}_{\bm{\Omega}} = \bm{\Gamma}\bar{\bm{A}}.
\end{equation}
Bring (\ref{eq Aomega}) into (\ref{eq cul_matrix}), we have,
\begin{equation}
\begin{split}
  \bm{C}_{\bm{\Omega}} &= \bm{\Gamma} \bar{\bm{A}} \bm{C_s} \bar{\bm{A}}^H \bm{\Gamma}^T, \\
  &= \bm{\Gamma} \bm{C} \bm{\Gamma}^T,
\end{split}
\end{equation}
where $\bm{C} = \bar{\bm{A}} \bm{C_s} \bar{\bm{A}}^H$ denotes the covariance matrix of the coarray output. It can be seen that compared to $\bm{C}_{\bm{\Omega}}$, $\bm{C}$ has a much larger dimensionality which can be used to increase the DoF.
To do this, let's first rewrite $\bm{C}$ as,
\begin{equation}
\begin{split}
 \bm{C} &= \sum_{k=1}^K c_{s_k} \left(\bar{\bm{a}}(\theta_k) \bar{\bm{a}}^H(\theta_k)\right) \\
 &\triangleq \sum_{k=1}^K c_{s_k} \bm{B}(\theta_k).
\end{split}
\end{equation}
And then define an atom set
\begin{equation}
 \mathcal{A} = \{ \bm{B}(\vartheta_k) = \bar{\bm{a}}(\vartheta_k) \bar{\bm{a}}^H(\vartheta_k), \vartheta_k \in (-90^{\circ},90^{\circ}] \}.
\end{equation}
Based on the atomic norm theory, the atomic norm of $\bm{C}$ can be defined as
\begin{equation}\label{eq atomic norm}
  \|\bm{C}\|_{\mathcal{A}} = \inf \left\{ \sum_k c_k | \bm{C} = \sum_k c_k \bm{B}(\theta_k), \bm{B}(\theta_k) \in \mathcal{A}, c_k > 0 \right\}.
\end{equation}
Although (\ref{eq atomic norm}) provides a proper decomposition of $\bm{C}$ by using the atoms in $\mathcal{A}$ with respect to the DOAs, it is still unknown how to compute the atomic norm from the definition. To solve this problem, the atomic norm can be transformed to a semidefinite programming (SDP). Formally, we have,
\begin{theorem}[\cite{zhouchengwei2018TSP}]
  The atomic norm defined in (\ref{eq atomic norm}) equals the optimal value of the following SDP,
  \begin{equation}
  \begin{split}
    &\|\bm{C}\|_{\mathcal{A}} = \min_{\bm{Z},\bm{u}}\ \frac{1}{2(2N+1)} \text{tr}[\bm{Z} + \bm{T}(\bm{u})] \\ &\text{s.t.}\ \begin{bmatrix}
  \bm{Z} & \bm{C} \\ \bm{C} & \bm{T}(\bm{u}) \end{bmatrix} \geq \bm{0},
  \end{split}
\end{equation}
where $\bm{T}(\bm{u}) \in \mathbb{C}^{(2N+1)\times (2N+1)}$ denotes a Toeplitz matrix with $\bm{u}$ being its first row.
\end{theorem}
It is shown that the DOAs are encoded in $\bm{T}(\bm{u})$ whose $(\bar{m},\bar{n})$-th element can be given as \cite{Tang2013offthegrid}
\begin{equation}
 \left(\bm{T}(\bm{u})\right)_{\bar{m},\bar{n}}= \sum_{k=1}^K c_{s_k} e^{ j2(m-n)\omega_k} \qquad m,n \in [N].
\end{equation}
Hence the DOAs can be retrieved from the Vandermonde decomposition of $\bm{T}(\bm{u})$ \cite{yangzai2015GLS}. Consequently, we propose the following atomic norm minimization (ANM) problem to retrieve $\bm{T}(\bm{u})$,\footnote{The term $\frac{1}{2(2N+1)}$ is omitted for brevity.}
\begin{equation}
\begin{split}
    &\min_{\bm{Z},\bm{u},\bm{C}}\ \text{tr}[\bm{Z} + \bm{T}(\bm{u})] \\
     \text{s.t.}\ &\begin{bmatrix}
  \bm{Z} & \bm{C} \\ \bm{C} & \bm{T}(\bm{u}) \end{bmatrix} \geq \bm{0},\\
   &\bm{\Gamma} \bm{C} \bm{\Gamma}^T = \bm{C}_{\bm{\Omega}}.
\end{split}
\end{equation}
In practice, since we can only observe the sample cumulant matrix of the sparse array output $\widehat{\bm{C}}_{\bm{\Omega}}$ rather than the exact one $\bm{C}_{\bm{\Omega}}$, the DOAs can be retrieved based on the following ANM problem,
\begin{equation}\label{eq convex problem}
\begin{split}
    &\min_{\bm{Z},\bm{u},\bm{C}}\ \text{tr}[\bm{Z} + \bm{T}(\bm{u})] \\
    \text{s.t.}\ &\begin{bmatrix}
  \bm{Z} & \bm{C} \\ \bm{C} & \bm{T}(\bm{u}) \end{bmatrix} \geq \bm{0},\\
   &\| \bm{C}_{\bm{\Omega}} - \widehat{\bm{C}}_{\bm{\Omega}} \|_F \leq \beta,
\end{split}
\end{equation}
where $\beta$ is an upper bound of the noise energy.

Solving the problem (\ref{eq convex problem}) by using CVX gives the estimation of $\bm{T}(\bm{u})$. Then the DOAs can be retrieved by using Vandermonde decomposition or the subspace methods such as root-MUSIC.


\begin{remark}
  By exploiting the coarray property of the redundancy array, our method is able to estimate more sources than sensors in terms of DOA estimation. In particular, the maximum number of detectable sources is $K_{\text{max}} = 2N$, which is greater than the number of sensors $2M+1$ if $M<N$.
\end{remark}

\subsection{Range Estimation of the NF Sources}

\begin{figure*}[!t]
\centerline{
\subfloat[DOA performance of FF sources]{\includegraphics[width=2.3in]{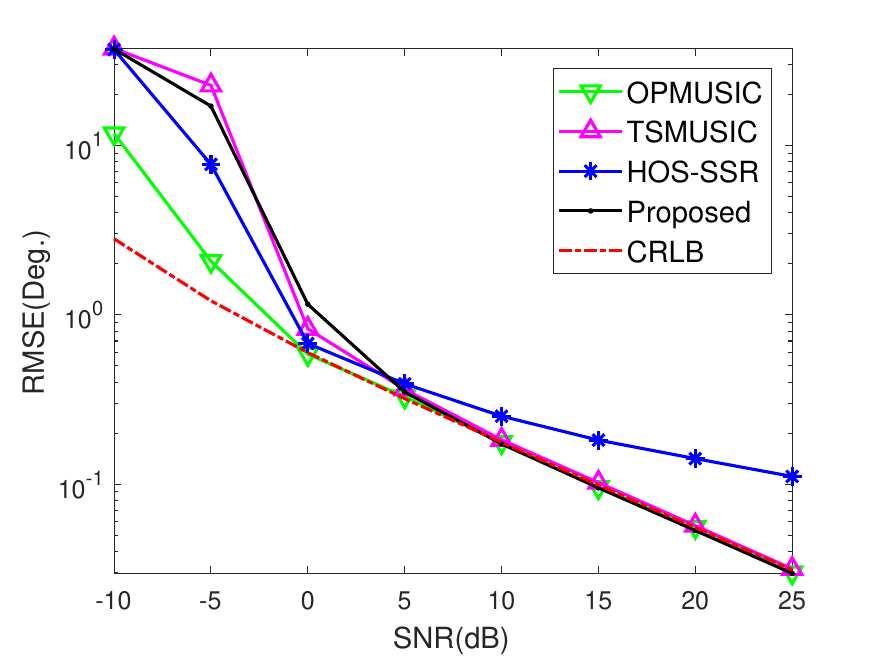}}
\hfil
\subfloat[DOA performance of NF sources]{\includegraphics[width=2.3in]{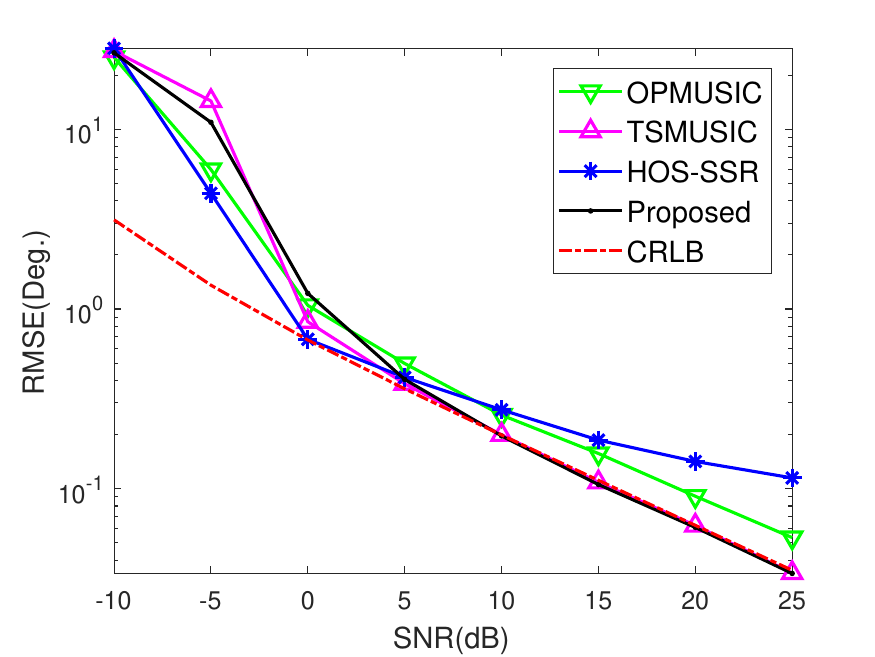}}
\hfil
\subfloat[Range performance of NF sources]{\includegraphics[width=2.3in]{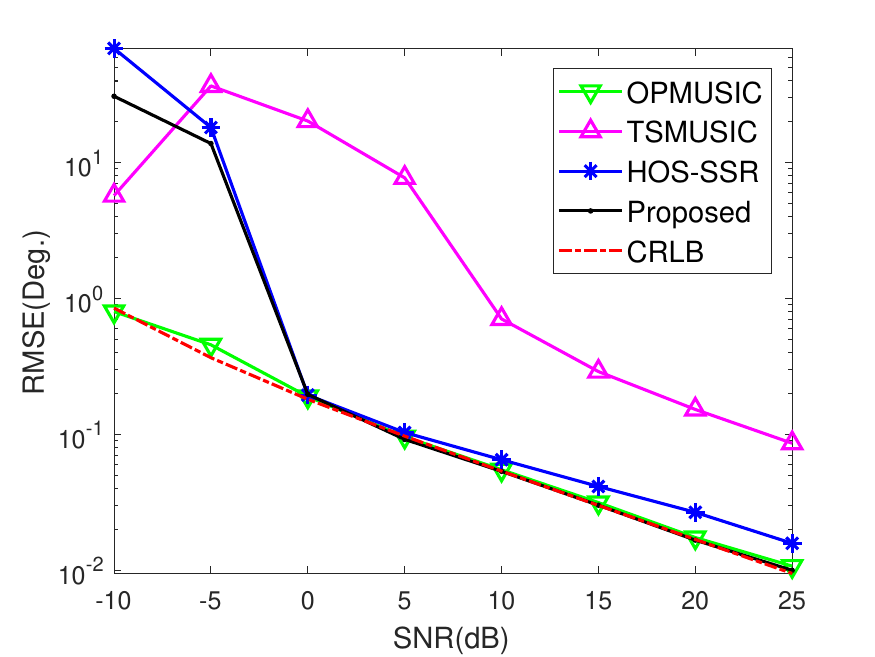}}}
\caption{DOA and range estimation comparison for one FF source from $\{0^{\circ},+\infty\}$ and one NF source from $\{25^{\circ},2\lambda\}$ impinging onto a $7$-element ULA with $L=200$.}
\label{fig RMSE SNR ULA}
\end{figure*}

\begin{figure*}[!t]
\centerline{
\subfloat[DOA performance of FF sources]{\includegraphics[width=2.3in]{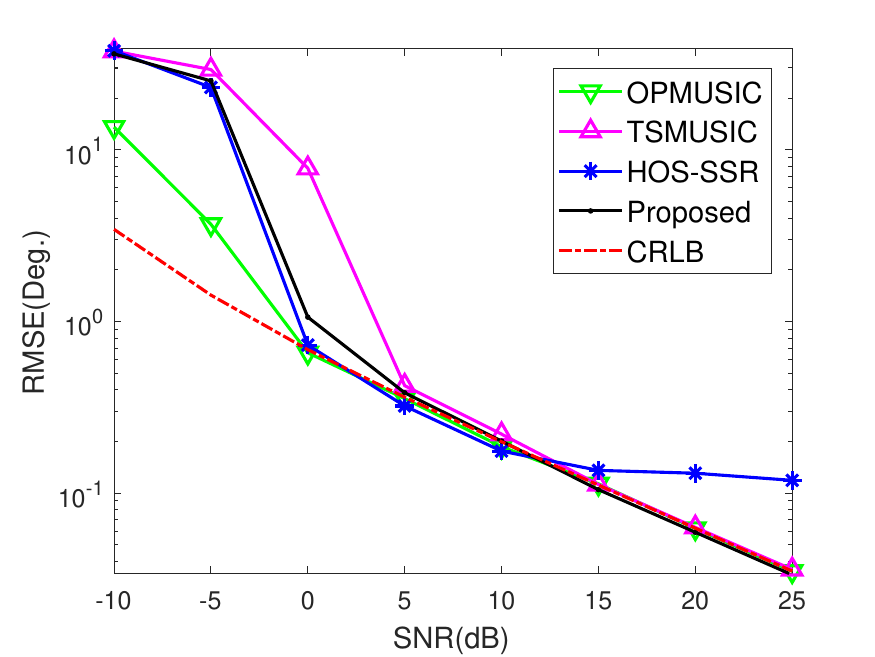}}
\hfil
\subfloat[DOA performance of NF sources]{\includegraphics[width=2.3in]{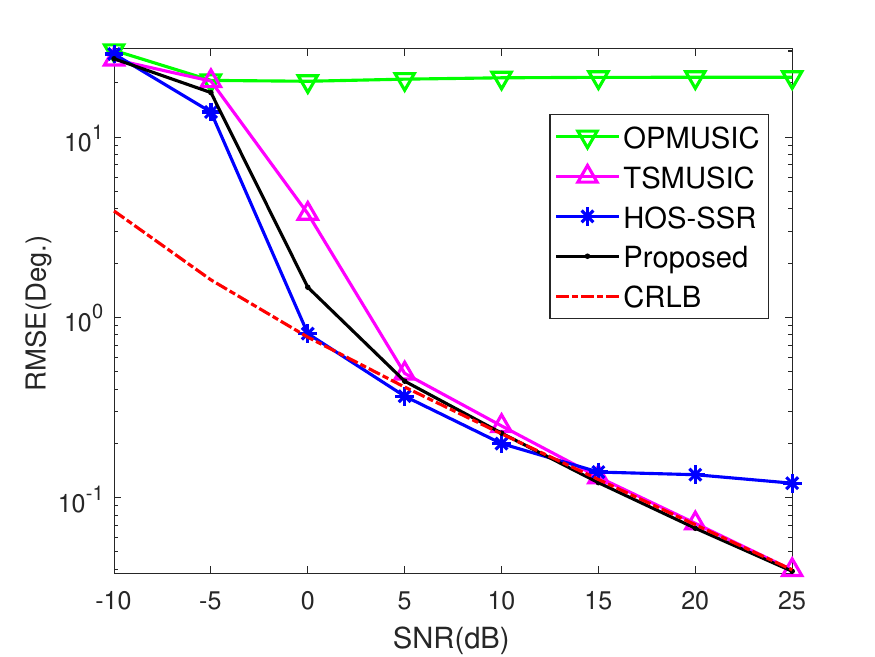}}
\hfil
\subfloat[Range performance of NF sources]{\includegraphics[width=2.3in]{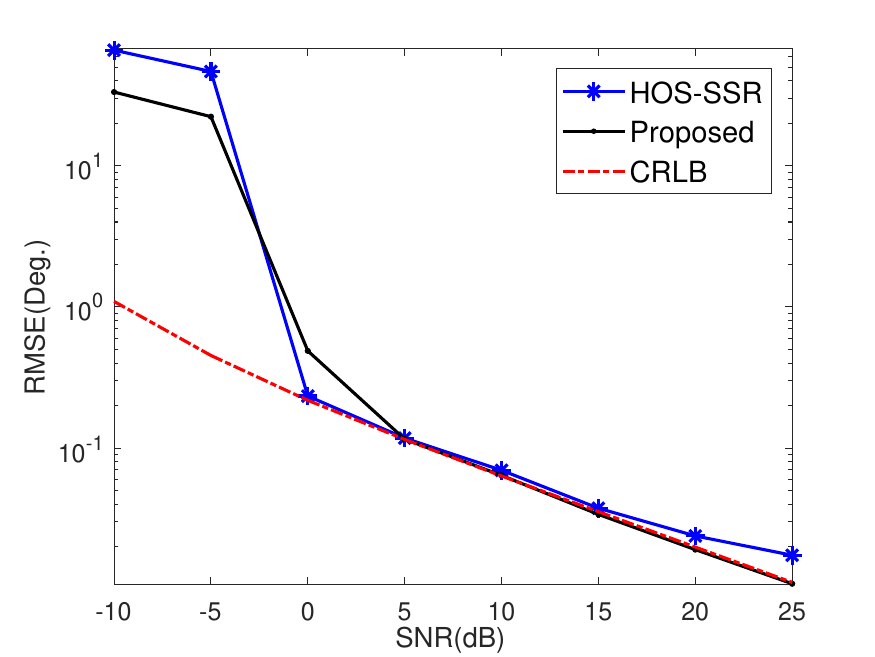}}}
\caption{DOA and range estimation comparison for one FF source from $\{0^{\circ},+\infty\}$ and one NF source from $\{25^{\circ},2\lambda\}$ impinging onto a $5$-element symmetric SLA with $L=200$.}
\label{fig RMSE SNR SLA}
\end{figure*}

In the previous part, we constructed a fourth-order cumulant matrix for DOA estimation based on the special structure of the array. In this part, however, it is unable to construct a fourth-order cumulant matrix for range estimation since the array is sparse. Hence, we utilize the covariance matrix of the array output and apply MUSIC-like algorithm to retrieve the ranges.
In particular, we first construct the sample covariance matrix $\widehat{\bm{R}}_{\bm{\Omega}}$ and then apply the eigen-decomposition to $\widehat{\bm{R}}_{\bm{\Omega}}$ to obtain the matrix $\bm{U}_{n}$ containing eigenvectors of $\widehat{\bm{R}}_{\bm{\Omega}}$ with respect to the $2M+1-K$ minimum eigenvalues. With the estimated DOAs $\hat{\theta}_k$, the range $r_k$ of the $k$-th signal can be obtained by solving the following problem,
\begin{equation}\label{eq range estimation}
  r_k = \min_{r} \ \bm{a}_{\bm{\Omega}}(\theta_k,r)\bm{U}_n \bm{U}_n^H \bm{a}^H_{\bm{\Omega}}(\theta_k,r),
\end{equation}
which can be solved by 1-D searching. Note that the FF source is a special NF source with $r\rightarrow +\infty$, we choose the range searching area as $(0.62(D^2/\lambda)^{0.5},r_{max} )$, where $r_{max}$ is selected to be larger than $2D^2/\lambda$ with $D$ being the array aperture. If the estimated range is located in the Fresnel region $(0.62(D^2/\lambda)^{0.5}, 2D^2/\lambda)$, the corresponding source is classified as the NF source. On the other hand, if the estimated range is larger than $2D^2/\lambda$, the source belongs to the FF source and we let $r=+\infty$. Note that the DOAs and ranges are automatically paired by using (\ref{eq range estimation}).

\section{Simulation Results}
\label{sec simulation}

In this section, we evaluate the performance of our proposed method with comparison to OPMUSIC \cite{He2012TSP_OPMUSIC}, TSMUSIC \cite{liang2010TSP_TSMUSIC}, high-order statistics based sparse signal representation method (HOS-SSR) \cite{Wang2012SPL_sparseNF} and the Crammer-Rao lower bound (CRLB) \cite{CRBforNF2005TSP}.\footnote{To reduce complexity, iterative grid refinement procedure is employed in HOS-SSR.}
The number of sources $K$ is assumed to be known for OPMUSIC, TSMUSIC and HOS-SSR but unknown for our method. Meanwhile, OPMUSIC requires the prior knowledge of the number of FF sources. 

We first consider a $7$-element ULA (i.e., $\bm{\Omega} = [3]$) with the sensor spacing $d = \lambda/4$ and assume two narrowband equal-power source signals consisting one FF source from $\{0^{\circ},+\infty\}$ and one NF source from $\{25^{\circ},2\lambda\}$ impinge onto the ULA. We set the number of snapshots $L = 200$ and evaluate these methods by comparing their RMSEs of the estimates with the SNR varying from $-10$dB to $25$dB and show the results in Fig. \ref{fig RMSE SNR ULA}. It can be observed from Fig. \ref{fig RMSE SNR ULA}(a) that our proposed method, OPMUSIC and TSMUSIC are able to approach and coincide with CRLB curve as the SNR grows. HOS-SSR deviates gradually from CRLB because of the model mismatch between the $\ell_0$-norm and $\ell_1$-norm minimization models. In Fig. \ref{fig RMSE SNR ULA}(b), our method and TSMUSIC can also show satisfying performance whereas OPMUSIC shows worse accuracy since it only utilizes partial information of the covariance matrix of the array output to estimate the DOAs of the NF sources. In Fig. \ref{fig RMSE SNR ULA}(c), we can see that TSMUSIC cannot provide good performance and there exists a large gap between the curves of TSMUSIC and CRLB. In contrast, OPMUSIC and our method can coincide with the CRLB and OPMUSIC performs better in low SNR region. HOS-SSR deviates from the CRLB in range estimation as well.

In the second experiment, we replace the ULA with a $5$-element symmetric SLA with $\bm{\Omega}=\{-3,-2,0,2,3\}$ and other settings are the same as Fig. \ref{fig RMSE SNR ULA}. From the RMSEs comparison displayed in Fig. \ref{fig RMSE SNR SLA} we can see that these four methods show similar performance as in the ULA case in DOA estimation of FF source. While for NF source localization, OPMUSIC requires spatial smoothing hence we can observe that it fails in Fig. \ref{fig RMSE SNR SLA}(b). For range estimation, TSMUSIC requires the uniform array structure and thus fails in the SLA case. Therefore, we omit the curves of OPMUSIC and TSMUSIC in Fig. \ref{fig RMSE SNR SLA}(c) from which it can be seen that our proposed method can still approach the CRLB while HOS-SSR deviates from the CRLB as the SNR grows.
In summary, our method performs better than other compared methods in DOA and range estimation in both ULA and SLA cases.

\section{Conclusions}
\label{conclusions}

In this paper, we proposed a mixed source localization based on ANM with symmetric redundancy linear arrays, including ULA, Cantor array, Fractal array and any symmetric redundancy SLAs. The proposed method does not require discretization of the angle space as well as any prior knowledge on the number of sources while still provides super-resolution in DOA and range estimation as compared to subspace and sparse methods in various scenarios.

\bibliographystyle{IEEEbib}

\end{document}